# Correlation between Foam-Bubble Size and Drag Coefficient in Hurricane Conditions

Ephim Golbraikh,[1] and Yuri M. Shtemler,[2]

[1]Department of Physics, Ben-Gurion University of the Negev, P.O. Box 653, Beer-Sheva 84105, Israel

[2]Department of Mechanical Engineering, Ben-Gurion University of the Negev, P.O. Box 653, Beer-Sheva 84105, Israel

**Abstract**

Recently proposed model of foam impact on the air–sea drag coefficient $C_d$ has been employed for the estimation of the efficient foam-bubble radius $R_b$ variation with wind speed $U_{10}$ in hurricane conditions. The model relates $C_d(U_{10})$ with the efficient roughness length $Z_{eff}(U_{10})$ represented as a sum of aerodynamic roughness lengths of the foam-free and foam-covered sea surfaces $Z_w(U_{10})$, and $Z_f(U_{10})$ weighted with the foam coverage coefficient $\alpha_f(U_{10})$. This relation is treated for known phenomenological distributions $C_d(U_{10})$, $Z_w(U_{10})$ and $\alpha_f(U_{10})$ at strong wind speeds as an inverse problem for the efficient roughness parameter of foam-covered sea surface $Z_f(U_{10})$.

The present work is motivated by recent experimental and theoretical studies of the momentum transfer from strong winds to the sea in hurricane conditions and by the progress in electromagnetic microwave scattering and measurements of brightness temperature of the air-sea interface (e.g., Powell et al., 2003; Donelan et al., 2004; Bye and Jenkins, 2006; Black et al, 2007; Jarosz et al., 2007; Shtemler et al., 2010; Holthuijsen et al., 2012; Soloviev et al., 2014; Golbraikh and Shtemler, 2016 and references therein).

The phenomenological parameterizations of the drag coefficient at low-to-hurricane wind speeds are commonly based on the assumption of a logarithmic law for vertical variation of the mean wind speed, $U$ $[ms^{-1}]$ (e.g., see details in Powell et al., 2003):

$$U/U_* = \kappa^{-1} \ln(Z/Z_{eff}), \qquad (1)$$

where $Z$ $[m]$ is the ordinate referred to the sea surface, $\kappa = 0.4$ is Karman's constant; $Z_{eff}[m]$ is the efficient roughness of sea surface; $U_*$ $[ms^{-1}]$ is the surface friction velocity related to the surface momentum flux $\tau = \rho U_*^2 = \rho C_d U_{10}^2$, where $\rho$ is the air density; $C_d$ is the drag coefficient, and $U_{10}$ $[ms^{-1}]$ is the wind speed at the reference height $L = 10[m]$. Then $C_d$ can be expressed using (1):

$$C_d = U_*^2/U_{10}^2 = \kappa^2/\ln^2(L/Z_{eff}). \tag{2}$$

The formula (2) can give either the value $C_d$ for a known $Z_{eff}$ or, vice versa, the value $Z_{eff}$ for a known $C_d$ both in the open-sea and laboratory conditions. Thus, an efficient roughness length determined by equation (2) corresponds to the averaging over the total (foam-free and foam-covered) underlying surface. However, such averaging has a minor physical meaning because of quite different properties of the foam-free and foam-covered sea surfaces, which are exhibited when they are probed by electromagnetic microwave scattering or their brightness temperature is measured (Wei, 2013; Guo et al., 2001; Reul and Chapron, 2003).

For the open-sea conditions, there are various methods determining the air-sea momentum transfer. Note one of them ("top-down" experimental technique) based on the measurements of $U$ at relatively large height $Z = H$, fitted by a least-squares line in the semi-log variable plane $\{U, \log H\}$, and then extrapolated to the sea level, where the mean wind speed $U = 0$ yields the value of $Z_{eff}$ (e.g. Powell et al., 2003; Black et al, 2007; Holthuijsen et al., 2012). Another evaluation of the air-sea momentum exchange can be achieved by measurements of the upper sea currents from the sea side of the air-sea interface ("bottom-up" experimental technique, Jarosz et al., 2007). The drag coefficient is contained in a simplified momentum balance equation, whose terms are estimated on the basis of the sea current observations. The authors argue that such technique generates a reliable and accurate determination of $C_d$ ($U_{10}$). The efficient drag coefficient $C_d$ ($U_{10}$) adopted from (Jarosz et al., 2007) and the roughness $Z_{eff}(U_{10})$ determined by the relation (2) are depicted by solid lines in Figs. 1a and 1b. Figures 1a and 1b also present the drag coefficient and roughness obtained by Powell et al., 2003; Holthuijsen et al., 2012. Note that at high wind velocities ($U_{10} > 33ms^{-1}$), the results obtained with "top-down" and "bottom-up" experimental techniques are in good agreement.

As it is well-known now, the behavior of $C_d$ is quite different in the open-sea and laboratory conditions under hurricane wind speeds $U_{10}$ higher than $33ms^{-1}$. Indeed, $C_d$ grows up to a saturation that occurs at $U_{10} \approx 33ms^{-1}$ in laboratory measurements (Donelan et al., 2004), but decreases after achieving the maximum value at $U_{10} \approx 33ms^{-1}$ in the open-sea measurements (Jarosz et al., 2007). As a result, this leads to totally different values of $Z_{eff}(U_{10})$ at hurricane wind speeds. As noted by Takagaki et al., 2016, the difference in $C_d$ behavior in field and laboratory experiments has not yet been fully clarified. The saturation of drag coefficient under hurricane conditions is frequently explained by the airflow separation from waves via sea drops and sprays (e.g., Donelan, 2004; Andreas and Emanuel, 2004; Andreas, 2004; Makin, 2005; Kudryavtsev, 2006; Kudryavtsev and Makin, 2011; Soloviev et al., 2014, etc.). In laboratory conditions, the foam coverage is formed by only a minor production of whitecaps due to wave breaking (see Fig. 2 in (Kandaurov et al., 2014) and Fig. 1 in (Takagaki et al., 2016)). It can be assumed that this occurs because of the finite length of the experimental high-speed wind-wave tank. Since the corresponding foam coverage is less than 20% of the interface, it was found to be insufficient to decrease $C_d$ at strong wind speeds, which results in the saturation of $C_d$. Most likely, this occurs due to the generation of droplet and spray effects, in contrast to the foam effect dominating in the open-sea conditions. To estimate the foam influence in laboratory measurements, a foam generator was installed near the outlet of the experimental wind-wave tank (Troitskaya et al., 2017). They noted a non-separated behavior of the air-sea flow in case of an enhanced surface roughness due to a stronger foam generation.

The present study is focused on the open-sea conditions. According to the scenario adopted below, the difference in the behavior of $C_d$ with $U_{10}$ in the field and laboratory measurements at strong winds occurs because of the difference in the foam coverage of the interface - almost foam-covered and almost foam-free water-surface under the field and laboratory conditions (Powell et al., 2003; Shtemler et al., 2010; Golbraikh and Shtemler, 2016). It is assumed that the results of laboratory experiments qualitatively correspond to the open-sea conditions with the foam-free interface. Following Golbraikh and Shtemler (2016) the efficient roughness length $Z_{eff}$ is represented as a sum of two aerodynamic roughness lengths for the foam-free and foam-covered surfaces $Z_w$ and $Z_f$ weighted by the fractional foam coverage, $\alpha_f$:

$$Z_{eff} = (1 - \alpha_f)Z_w + \alpha_f Z_f, \qquad (3)$$

where $\alpha_f = S_f/S$; $S = S_w + S_f$, $S_w$ and $S_f$ are the total foam-free and foam-covered areas, respectively. Later, the similar approach had been applied to surf zone study in (Macmahan, 2017).

In general, formulas (2) and (3) assume the knowledge of phenomenological distributions $Z_w(U_{10})$, $Z_f(U_{10})$ and $\alpha_f(U_{10})$ in the whole range of the low-to-hurricane wind speeds $U_{10}$. For open-sea conditions, foam coverage $\alpha_f$ vs. $U_{10}$ was adopted from the observation data of Holthuijsen et al. (2012) and approximated by

$$\alpha_f = \gamma \tanh\left[\alpha \exp\left(\beta \frac{U_{10}}{U_{10}^{(S)}}\right)\right], \tag{4}$$

where $\alpha = 0.00255$, $\beta \approx 8$, $\gamma = 0.98$, and the saturation velocity $U_{10}^{(S)} \approx 48\ [ms^{-1}]$ for hurricane conditions in the open-sea case.

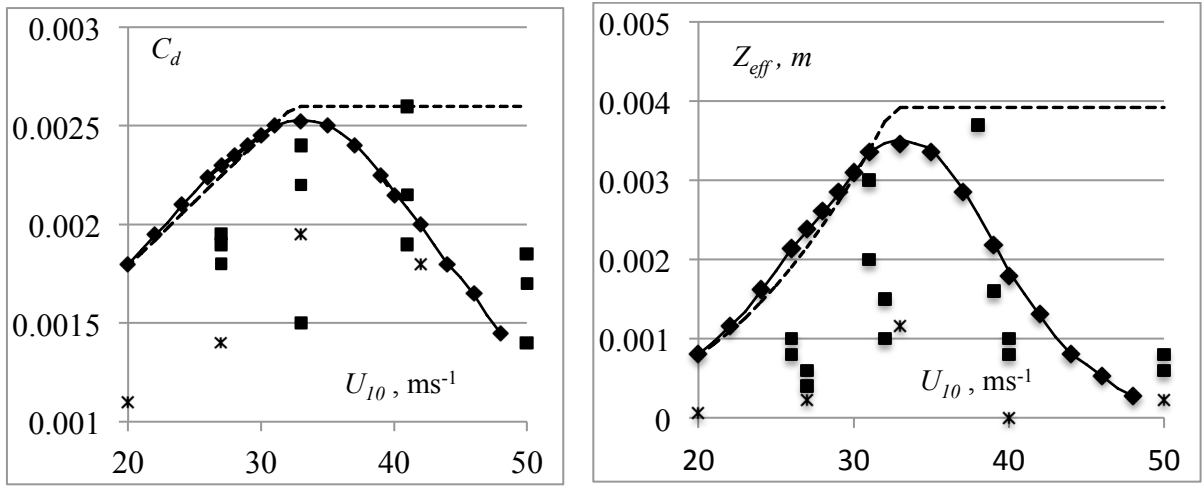

Figure 1. Efficient (a) drag coefficient $C_d$ and (b) roughness $Z_{eff}$ vs. $U_{10}$.

Solid lines interpolate the data for open-sea conditions (Jarosz et al., 2007).

Dashed lines depict the model data (2), (5) for foam-free conditions $(C_d = C_w, Z_{eff} = Z_w)$.

The field data: squares (Powell et al., 2003); diamonds (Jarosz et al., 2007);

crosses (Holthuijsen et al., 2012).

In a recent paper (Golbraikh and Shtemler, 2016), the authors defined $Z_w$ for field experiments by using a modified Charnock model (Charnock, 1955; Fairall et al., 2003; Edson et al., 2013), while the efficient roughness length $Z_f$ was assumed to be correlated with the characteristic size of the foam bubbles $R_b$ (by analogy with fixed beds). However, the field measurements of $R_b$ are rather scarce, and $Z_f(R_b)$ can be only estimated. For further simplicity, $Z_f$ was assumed to be a constant averaged value of the efficient radius of the foam bubble $Z_f = R_b \approx 0.3\ mm$. Then the substitution of these values of $Z_w$, $Z_f$ and $\alpha_f$ into (2) and (3) yields the efficient drag coefficient and roughness length. Although the resulting dependences $C_d(U_{10})$ and $Z_{eff}(U_{10})$ obtained by Golbraikh and Shtemler (2016) are in a fair agreement with available measurement data, their modeling has a rather over-simplified illustrative character. Namely, the modified Charnock model for $Z_w(U_{10})$ lost its accuracy at strong wind speeds, while the efficient (size-averaged) bubble radius $R_b$ and, equivalently, $Z_f$ should be varied with wind speed and may strongly affect the foam emissivity in hurricane conditions.

In the present work, a qualitative similarity is assumed between laboratory observations with small foam coverage and virtual field experiments with a foam-free interface. Thus, $Z_w(U_{10})$ is determined with the help of Large and Pond model (1981) for calculating $C_d(U_{10})$ in the open-sea conditions instead of modified Charnock model explored by Golbraikh and Shtemler, 2016. The Large-Pond model for the roughness of the interface $Z_w(U_{10})$ applied at $U_{10} < 26\ ms^{-1}$ in the open-sea conditions is modified below in accordance with the laboratory observations by Donelan et al., 2004. According to their laboratory observations, the drag coefficient $C_d$ is constant at $U_{10} > 33 ms^{-1}$ (at which the maximum value of $C_d$ in the field measurements is located). The latter condition yields the saturation value $C_d \cdot 10^3 = 2.6$, and the modified Large-Pond model is:

$$C_d \cdot 10^3 = \begin{cases} 1.14 & 4\ ms^{-1} < U_{10} \leq 10\ ms^{-1}, \\ 0.49 + 0.065 U_{10} & 10\ ms^{-1} < U_{10} \leq 33\ ms^{-1}, \\ 2.6 & 33\ ms^{-1} < U_{10}. \end{cases} \qquad (5)$$

Then the model (5) together with formula (2) yield $C_d = C_w, Z_{eff} = Z_w$ for the open-sea conditions, which are presented by dashed lines in Figures 1a and 1b.

In the present work, we also assume that the bubble size should be varied with wind speed. Unfortunately, no measurements have been done until now that evaluate the dependence of the foam-bubble size on wind speed (Reul and Chapron, 2003). The present study is aimed at solving an inverse problem which gives, using (3), an efficient foam roughness length $Z_f$ from the measurement data for the drag coefficient, which, in turn, determines the distribution $Z_{eff}(U_{10})$

$$Z_f = [Z_{eff} - (1 - \alpha_f) Z_w]/\alpha_f. \qquad (6)$$

Formula (6) determines $Z_f(U_{10})$ if the dependences $Z_w(U_{10})$ and $\alpha_f$ are given by (2), (4)-(5), while the dependence $Z_{eff}(U_{10})$ is specified by data obtained according to (Jarosz et al., 2007).

As can be seen from Fig. 2, the roughness length for the foam-covered sea-surface, $Z_f$, grows with wind velocity, $U_{10}$, up to the maximum value of $Z_f \approx 3.5\ mm$ at $U_{10} \simeq 26\ ms^{-1}$, it will remain at a near-constant value within the interval $U_{10} \simeq 26 \div 34\ ms^{-1}$ and will decreases with further growth of $U_{10}$.

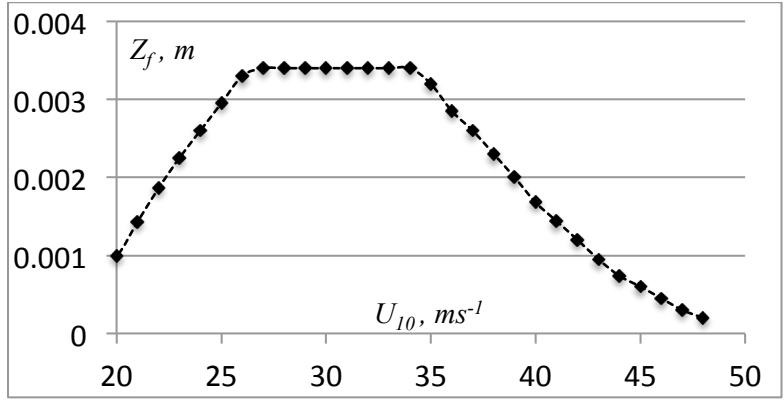

Figure 2. $Z_f$ vs. $U_{10}$ for the open-sea conditions (Jarosz et al., 2007), using the model (2), (4)-(6).

The wave breaking generates whitecap bubbles with a wide spectrum of their efficient radii typically ranging in size from tenths of a millimeter (short-living large bubbles) to several hundred microns (long-living small bubbles $0.2 \div 0.4\, mm$ that form streaks (Anguelova and Webster, 2006; Callaghan et al., 2008; Holthuijsen et al., 2012; Deike et al., 2016 and references therein). Assuming that the efficient length of the foam roughness in Fig. 2 correlates with the behavior of the efficient radii of whitecaps and streaks in the regions of their domination (Holthuijsen et al., 2012): whitecaps domination at low wind speeds ($U_{10} \lesssim 25\, ms^{-1}$), streaks domination at high wind speeds ($U_{10} \gtrsim 35\, ms^{-1}$), and coexistence of whitecaps and streaks at intermediate wind speeds ($25\, ms^{-1} \lesssim U_{10} \lesssim 35\, ms^{-1}$).

Thus, an inverse problem solved herein provides a phenomenological distribution for the efficient roughness of the foam-covered sea surface $Z_f(U_{10})$ correlated with the efficient foam-bubble radius $R_b(U_{10})$ in the region of a hurricane wind speed. We believe that the present study is the first step toward a reliable solution of the direct problem (2)-(3) proposed by Golbraikh and Shtemler (2016) for phenomenological distributions $C_d(U_{10})$ and $Z_{eff}(U_{10})$. The direct problem solution provides an alternative to commonly accepted models (1)-(2) for the estimation of drag coefficient and roughness. The direct problem (2)-(3) does not require measurements of either air wind or sea-water current in hurricane conditions. This approach is based on the correlation between the efficient aerodynamic roughness of the foam-covered sea surface $Z_f$ and the response to electromagnetic microwave scattering of the air-sea surface.

The direct problem (2)-(3) is based on the knowledge of phenomenological distributions $\alpha_f(U_{10})$, $Z_w(U_{10})$ and $Z_f(U_{10})$. The observation data for $\alpha_f(U_{10})$ are well established now, the roughness length of the foam-free sea surface $Z_w(U_{10})$ can be provided (as proposed above) by the model (2), (5) based on the laboratory experimental data, while reliable experimental data for the roughness length of the foam-covered sea surface $Z_f(U_{10})$ are not available yet. The latter forced us to confine ourselves by solving the inverse problem (2)-(3) that predicts the values of $Z_f(U_{10})$ assuming that $C_d(U_{10})$ is known from the conventional models (1)-(2).

Note that this approach can possible be further developed by introducing the whitecap- and streak- roughness lengths separately as , $Z_{fw}$ and $Z_{fs}$, in the expression for $Z_{eff}$, where $Z_{eff} = (1-\alpha_f)Z_w + \alpha_{fw}Z_{fw} + \alpha_{fs}Z_{fs}, \quad (\alpha_f = \alpha_{fw} + \alpha_{fs})$, (7) instead of the foam roughness length $Z_f$ in (3) when averaged over large and small bubbles of the whitecaps and streaks. Behavior of $\alpha_{fw}$ and $\alpha_{fs}$ were obtained in (Holthuijsen et al., 2012): $\alpha_{fw}$ increases with wind speed up to its saturation value $\approx 0.05$ at 24 $ms^{-1}$; $\alpha_{fs}$ increases up to the value $\approx 0.95$ at wind speeds lager than $42\ ms^{-1}$. However, this improvement is left to future study when measurement data for $Z_{fw}$ and $Z_{fs}$ will be available in the whole region of wind speeds.